\documentclass[twocolumn,showpacs,preprintnumbers,amsmath,amssymb,usletter]{revtex4}

\usepackage{amsmath,amssymb,bbm,epsfig}
\usepackage[english]{babel}
\newcommand{\be}{\begin{equation}}
\newcommand{\ee}{\end{equation}}
\newcommand{\ba}{\begin{array}}
\newcommand{\ea}{\end{array}}
\newcommand{\bqa}{\begin{eqnarray}}
\newcommand{\eqa}{\end{eqnarray}}

\newcommand{\um}{\mathbbm{1}}
\newcommand{\tr}{\mbox{Tr}}

\newcommand{\bra}[1]{\ensuremath{\langle #1 |}}
\newcommand{\ket}[1]{\ensuremath{| #1 \rangle}}

\newcommand{\ovl}[2]{\ensuremath{\langle #1 | #2 \rangle}}

\begin{document}

\title{Smooth optimal control with Floquet theory}

\author{Bj\"{o}rn Bartels, Florian Mintert}

\affiliation{Freiburg Institute for Advanced Studies, Albert-Ludwigs University of Freiburg, Albertstr. 19, 79104 Freiburg, Germany}

\date{\today}

\begin{abstract}
This paper describes an approach to construct temporally shaped control pulses that drive a quantum system towards desired properties.
A parametrization in terms of periodic functions with pre-defined frequencies permits to realize a smooth, typically simple shape of the pulses;
their optimization can be performed based on a variational analysis with Floquet theory.
As we show with selected specific examples, this approach permits to control the dynamics of interacting spins, such that gate operations and entanglement dynamics can be implemented with very high accuracy.
\end{abstract}

\pacs{
42.50.Dv, 
03.65.Yz, 
02.30.Yy}

\maketitle

Research on quantum mechanical systems is currently undergoing a process of substantial changes:
whereas in the last decades the effort was mostly on the observation and description of quantum mechanical properties,
currently the option to manipulate and control quantum systems is moving into focus.
On the one hand, this is due to the technological advances that permit isolating quantum systems {\it e.g.} in ion traps \cite{blatt_physrep,nature:453:1008} or optical lattices \cite{lattice_review},
manipulating them coherently
and letting quantum systems of different kinds interact with each other \cite{L.Childress10132006,Bleszynski-Jayich09102009}.
On the other hand, there is the prospect to exploit intrinsically quantum mechanical properties to engineer devices with performance characteristics far beyond the classically achievable.
Whereas secure communication \cite{zugspitze}
or teleportation \cite{Marcikic:2003fk} based on quantum protocols is well established by now and
the possibility to simulate quantum mechanical many-body systems with quantum simulators \cite{natphys_4_757,Struck19082011}
is a growing field,
new perspectives to exploit quantum mechanical coherence phenomena, for example in energy provision \cite{engel,nature:463:644},
are just emerging.

Beyond the experimental capability to control quantum systems, any type of {\it quantum engineering} also needs appropriate theoretical tools that allow an experimentalist to extract optimal performance under given limitations of control, as typically imposed by power or frequency range of driving fields.
Optimal control theory \cite{dalessandro,krotov,Khaneja2005296,1367-2630-12-7-075008}
provides elaborate and efficient schemes to identify {\it e.g.} shapes of laser or microwave pulses that drive a system towards desired properties.
A particularly astonishing property of pulses designed by optimal control techniques is their robustness against experimental imperfections \cite{Schneider2007291,PhysRevA.65.021403};
for example inhomogeneous broadening in ensembles of quantum systems can be compensated essentially completely through suitably designed control pulses \cite{PhysRevA.39.5725,Abramovich199330,skinner}. 

A disadvantage of these pulses is that they typically contain many frequency components;
besides potential experimental challenges to generate such pulses,
the complicated structure of these pulses renders it essentially impossible to {\em understand} why they result in their astonishing performance.
In particular if we want to push the envelope to large many-body systems, the answer to the question of `{\em why}' will become more and more important rather than the observation `{\em that}' one can identify suitable pulses.
The aim here is therefore to strive for an approach that permits to limit a pulse to given frequency components \cite{Skinner2010248,PhysRevA.84.022326,moore:134113,PhysRevA.76.052304}, resulting in rather simple pulse shapes. 

The typical scenario of optimal control theory is dynamics induced by a system- or drift-Hamiltonian ${\cal H}_D$ and an often time-dependent control Hamiltonian ${\cal H}_C(t)$.
The system Hamiltonian is assumed to be inaccessible to control, {\it i.e.} it contains no adjustable components, whereas the control Hamiltonian can be tailored in a time-dependent fashion. 
The central question of interest is to determine how the propagator ${\cal U}$
induced by the time-dependent Hamiltonian
${\cal H}(t)={\cal H}_D+{\cal H}_C(t)$ changes if the control Hamiltonian is varied,
since this provides the basis for an iterative improvement of ${\cal H}_C(t)$ \cite{krotov,Khaneja2005296}.
Typically, the time-dependence of the control Hamiltonian stems from the time-dependence of an externally applied control field, like a laser- or microwave field,
so that ${\cal H}_C(t)=\sum_if_i(t)\mathbf{h}_i$,
where the time-independent Hamiltonians $\mathbf{h}_i$ describe the coupling of the system to the various control fields, and $f_i(t)$ is the time-dependence of the amplitude of the corresponding field.
The objects of interest thus are the functional derivatives $\delta{\cal U}[f_i]/\delta f_i$. 

In frequently employed control algorithms like Krotov \cite{krotov} and GRAPE \cite{Khaneja2005296,steffen2ndorder} the time interval in which control is exerted is divided into many short sub-intervals and the time-dependence of control fields is described by step-like pulses that are constant within each sub-interval.
Consequently, the time-evolution operator ${\cal U}$ becomes a product of propagators for the individual sub-intervals that are induced by time-independent Hamiltonians,
and the functional derivatives reduce to ordinary derivatives with respect to the fields' amplitudes in the individual sub-intervals.
These ordinary derivatives can be obtained rather efficiently, which results in the astonishing performance of Krotov and GRAPE.
This decomposition of the control fields' time-dependence in piecewise constant segments, however, often results in pulse-shapes that contain high-frequency components \cite{skinner}.

In this work, we will follow a parametrization of the control pulse in terms of Fourier modes \cite{PhysRevA.76.052304,PhysRevLett.106.190501}. By doing this, high-frequency components can be excluded by construction. We describe how the desired derivatives with respect to the control parameters can be obtained with Floquet theory,
and demonstrate the performance of this approach for the control of spin interactions.
In particular, the specific examples in Sec. \ref{examples} show that pulses with very few frequency components (typically less than ten) are sufficient to induce time-optimal gates or control entanglement dynamics with very high accuracy.

\section{Method}

In the following, we employ an expansion of the field amplitudes
\be
f_i(t)=\sum_j a_{ij}\ g_{j}(t)
\label{eq:para}
\ee
in periodic functions $g_j(t)$ with a given periodicity $T$.
Since the period $T$ can be chosen to coincide with the duration over which control is exerted, the assumption of periodicity does not result in any restriction in practice,
but due to this periodicity Floquet theory \cite{floquet,PhysRevA.39.5725,Skinner2010248} can be invoked to determine the propagator ${\cal U}$.
According to Floquet's theorem, the Schr\"odinger equation with a periodically time-dependent Hamiltonian with periodicity $T$
has a complete set of solutions
\be
\ket{\Psi_k(t)}=e^{-i\varepsilon_k t}\ket{\Phi_k(t)}\label{eq:solutions}
\ee
with the quasi-energies $\varepsilon_k$ and periodically time-dependent state vectors $\ket{\Phi_k(t)}=\ket{\Phi_k(t+T)}$,
which are the eigenvalues and corresponding eigenstates of the Floquet operator
${\cal K}={\cal H}-i\partial_t$.

Due to their periodicity, the state vectors $\ket{\Phi_k(t)}$ can be decomposed in a discrete Fourier series
$\ket{\Phi_k(t)}=\sum_\nu\ket{\Phi_{k\nu}}e^{i\nu\Omega t}$ with $\Omega=2\pi/T$.
Introducing a state vector $\ket{\nu}$ for the function $e^{i\nu\Omega t}$
allows to express the time-derivative $\partial_t$

through the number operator $\hat N\ket{\nu}=\nu\ket{\nu}$,
and the Floquet operator admits a matrix representation
\be
{\cal K}=\sum_\nu{\cal H}_\nu \otimes\pi_\nu+\um\otimes\Omega \hat N\ ,\label{eq:floq}
\ee
with raising operators $\pi_\nu\ket{\mu}=\ket{\nu+\mu}$ and Fourier components $\mathcal{H}_\nu=1/T\int_0^T\mathrm dt\,e^{-i\nu\Omega t}\mathcal{H}(t)$ of the Hamiltonian $\mathcal{H}(t)$.

${\cal K}$ has a $2\pi/T$-periodic spectrum, and the eigenvalues $\varepsilon_k$ and eigenvectors $\ket{\chi_k}$ within one Brillouin zone, {\it i.e.} an interval of width $2\pi/T$, determine a complete set of solutions
\be
\ket{\Psi_k(t)}=\sum_\nu \ovl{\nu}{\chi_k}e^{i(\nu\Omega-\varepsilon_k)t}\
\label{eq:psift}
\ee
of the Schr\"odinger equation.

Together with the solutions given in Eq. \eqref{eq:solutions} the propagator 
reads
\be
{\cal U}(t)=\sum_k\ket{\Psi_k(t)}\bra{\Psi_k(0)}=\sum_k e^{-i\varepsilon_k t}\ket{\Phi_k(t)}\bra{\Phi_k(0)}\ .
\label{eq:propagator}
\ee

Since ${\cal U}$ is determined completely in terms of the eigensystem of ${\cal K}$ via Eqs. \eqref{eq:propagator} and \eqref{eq:psift},
the desired derivatives of ${\cal U}$ with respect to the control parameters $a_{ij}$ are given by the corresponding derivatives of $\varepsilon_k$ and $\ket{\chi_k}$;
these, in turn, are readily obtained from a perturbative analysis of ${\cal K}$.
As ${\cal K}$ depends linearly on the control parameters $a_{ij}$,
the linear expansion
\be
{\cal K}=
{\cal K}_0+\sum_{ij}{\cal K}_{ij}(a_{ij}-a_{ij}^{(0)})
\ee
with ${\cal K}_0={\cal K}|_{a_{ij}=a_{ij}^{(0)}}$ and ${\cal K}_{ij}=\partial_{a_{ij}}{\cal K}$ is exact for any choice of $a_{ij}^{(0)}$.
The dependence of $\varepsilon_k$ and $\ket{\chi_k}$ on the control parameters $a_{ij}$ can therefore be described by a series expansion with the small parameters $a_{ij}-a_{ij}^{(0)}$,
and the expansion coefficients in $n-$th order are given by the $n-$th order perturbative corrections with ${\cal K}_0$ as unperturbed operator and the perturbations ${\cal K}_{ij}$.
The explicit form of the derivatives reads
\bqa
\frac{\partial\varepsilon_k}{\partial a_{ij}}&=&\bra{\chi_k}{\cal K}_{ij}\ket{\chi_k}\ ,\
\frac{\partial\ket{\chi_k}}{\partial a_{ij}}=-{\cal I}_{k}{\cal T}_{kij}\ket{\chi_k}\ ,\nonumber\\
\frac{\partial^2\varepsilon_k}{\partial a_{ij}\partial a_{pq}}&=&-\bra{\chi_k}\left({\cal T}_{kij}{\cal I}_{k}{\cal T}_{kpq}+{\cal T}_{kpq}{\cal I}_{k}{\cal T}_{kij}\right)\ket{\chi_k} \ ,\label{eq:pert}\\
\frac{\partial^2\ket{\chi_k}}{\partial a_{ij}\partial a_{pq}}&=&
\left(\ {\cal I}_{k}{\cal T}_{kij}{\cal I}_{k}{\cal T}_{kpq}+\ {\cal I}_{k}{\cal T}_{kpq}{\cal I}_{k}{\cal T}_{kij}\right)\ket{\chi_k}\nonumber\\
& &-\frac{1}{2}\bra{\chi_k}\left({\cal T}_{kij}{\cal I}_{k}^2{\cal T}_{kpq}+{\cal T}_{kpq}{\cal I}_{k}^2{\cal T}_{kij}\right)\ket{\chi_k}\ket{\chi_k}\nonumber
\eqa
where $\ket{\chi_k}$ is the eigenvector of ${\cal K}_0$ to the eigenvalue $\varepsilon_k$, ${\cal I}_{k}$ is the pseudo-inverse of ${\cal K}_0-\varepsilon_k$, {\it i.e.} the inverse restricted to the sub-space orthogonal to $\ket{\chi_k}$, and
${\cal T}_{kij}={\cal K}_{ij}-\frac{\partial \varepsilon_k}{\partial a_{ij}}\mathbbm{1}$.

These derivatives determine the derivatives of the propagator ${\cal U}$, and, in turn, of any target functional, {\it i.e.} property of the final state $\varrho(t_f)={\cal U}(t_f)\varrho(0){\cal U}^\dagger(t_f)$ at a given final time $t_f$ that is to be optimized.
Similar to Krotov and GRAPE \cite{krotov,Khaneja2005296}, this information can be used to improve a pulse sequence as parametrized by the $a_{ij}$,
and close-to-optimal pulses are obtained in an iterative manner.

The main goal of the present approach was to arrive at control pulses with a potentially narrow spectral range.
Yet, the above expansion of a control sequence also yields various additional benefits:
\begin{itemize}

\item[-]A control pulse is parametrized with comparatively few parameters; as shown below, pulses with less than 10 frequency components yield very satisfactory results, whereas conventional methods with piecewise constant functions typically require the optimization of several hundred or thousand parameters \cite{Schneider2007291,PhysRevA.65.021403}.
\item[-]Also higher order derivatives of ${\cal U}$ with respect to the control fields can be obtained exactly, what improves convergence as compared to algorithms based on first derivatives only \cite{steffen2ndorder,PhysRevLett.108.110504}.
The curvature is described by quadratically more scalars than the gradient, which makes its evaluation intrinsically more expensive, but due to the small number of control parameters in the present framework,
accessing the exact curvature is certainly practical.
If, however, this effort is to be avoided, the (scalar) curvature along a specific direction (typically the gradient) is also available in a perturbative analysis with a single perturbation $\sum_{ij}b_{ij}{\cal K}_{ij}$, where $b_{ij}$ determines the direction along which the curvature is to be determined.
\item[-]Since the complete time-propagation operator is constructed, as opposed to its action on a pre-defined initial state, accessing non-linear target functionals \cite{PhysRevLett.105.020501} hardly requires additional effort as compared to typical linear functionals like fidelity.
\item[-]Boundary conditions, such as smooth switching on and off of the pulses, can be realized easily through a suitable parametrization in Eq. \eqref{eq:para}.
\item[-]In order to find pulses which achieve a certain control goal in minimal time, we can use the duration of the pulse as a real control {\it parameter}, such that the algorithm searches for optimal solutions in the extended space of field amplitudes $a_{ij}$ and pulse duration $t_f$.
\item[-]Since all dynamics are given by time-dependent exponential functions, which can be integrated analytically, time-averaged target functionals can also be implemented without additional overhead.
This permits, for example, to minimize the deviation from a desired dynamics over a finite time interval.
\end{itemize}

\section{Examples}\label{examples}

Having established our method, we will demonstrate its versatility by applying it to three selected problems from quantum information theory.

\subsection{Entangling gates in minimal time}\label{minentang}

A fundamental task in quantum computation is to construct gates that entangle two spins \cite{PhysRevA.84.042315,PhysRevLett.111.050503,PhysRevLett.109.060501}. The unavoidable presence of decoherence makes it necessary to perform this operation in the shortest possible time. The easiest way to obtain time-optimal pulses is to let the algorithm find a solution with maximal entanglement for an initial pulse duration $t_f=t_1$ and then repeat for successively smaller times $t_k<t_1$ until a certain threshold of entanglement cannot be reached anymore \cite{0953-4075-44-15-154011}. This method can be simplified significantly in the following way: Instead of iterating many times for different pulse durations, in the framework of our method, the duration $t_f$ of the control pulse can be introduced as an additional control parameter, such that only a single optimization run is needed.

In our framework, as we want the control field to be switched on at $t=0$ and switched off at $t=t_f$ ($f_i(0)=f_i(t_f)=0$), we have to ensure that the zeros of the lowest frequency component $\Omega$ coincide with the begin and the end of the control interval. This means that the product $\Omega t_f$ has to be kept fixed, so that decreasing the pulse duration $t_f$ implies an increase of the fundamental frequency $\Omega$. If one takes trigonometric functions as the basis $g_j(t)$ of the pulse, as we will do in this paper, one has $\Omega t_f=\pi$. In order to incorporate this constraint, one eliminates $t_f$ in Eq. \eqref{eq:propagator} by $t_f=\pi/\Omega$. The fact that $\Omega$ is now used as an additional control parameter entails the following advantage: Since the Floquet operator \eqref{eq:floq} is linear in $\Omega$, the derivatives of the propagator $\mathcal{U}$ with respect to $\Omega$ can be obtained by using the same perturbation ansatz \eqref{eq:pert} as for the control amplitudes $a_{ij}$.

For the problem of creating time-optimal gates $\mathcal{U}_d$, the target functional reads $\mathcal{F}=\mathcal{F}_0-\mathcal{F}_p$. Here, $\mathcal{F}_0=\text{Re}(\tr(\mathcal{U}(t_f)^\dagger\mathcal{U}_d))/2$ is the fidelity of the implemented gate, while $\mathcal{F}_p$ penalizes long pulse durations. The simplest choice for the penalty functional is $\mathcal{F}_p=pt_f$, where $p\geq 0$ is a parameter that decides on how strongly long pulse durations are penalized. If one chooses a strong penalty right from the beginning of the optimization, the algorithm tends to implement a fast gate but with very mediocre fidelity. For this reason, we started our algorithm with $p=0$ and successively increased $p$, as soon as $\mathcal{F}_0$ exceeds a certain fidelity threshold $\mathcal{F}_\text{thr}$.

In the following, we consider the system Hamiltonian
\be
\mathcal{H}_0=\frac{\omega_1}{2}\sigma_z^{(1)}+\frac{\omega_2}{2}\sigma_z^{(2)}+g_x\sigma_x^{(1)}\sigma_x^{(2)}+g_y\sigma_y^{(1)}\sigma_y^{(2)}\label{ham2s}
\ee
with xx/yy-couplings of the spins and a (possibly small) splitting in $z$-direction, chosen at random. We assume control over the $x$- and $y$-component of both spins by the control Hamiltonian
\be
\mathcal{H}_c(t)=\sum_{\substack{k=x,y\\j=1,2}}f_k^{(j)}(t)\sigma_k^{(j)},\label{ham2sc}
\ee
where the control pulse
\be
 f_k^{(j)}(t)=\sum_{n=1}^{n_\text{max}}a_n^{(k,j)}\sin(n\Omega t)\label{ham2sp}
\ee
contains $n_\text{max}$ frequency components of a fundamental frequency $\Omega$.

\begin{figure}[ht]
\includegraphics[width=0.4\textwidth,angle=0]{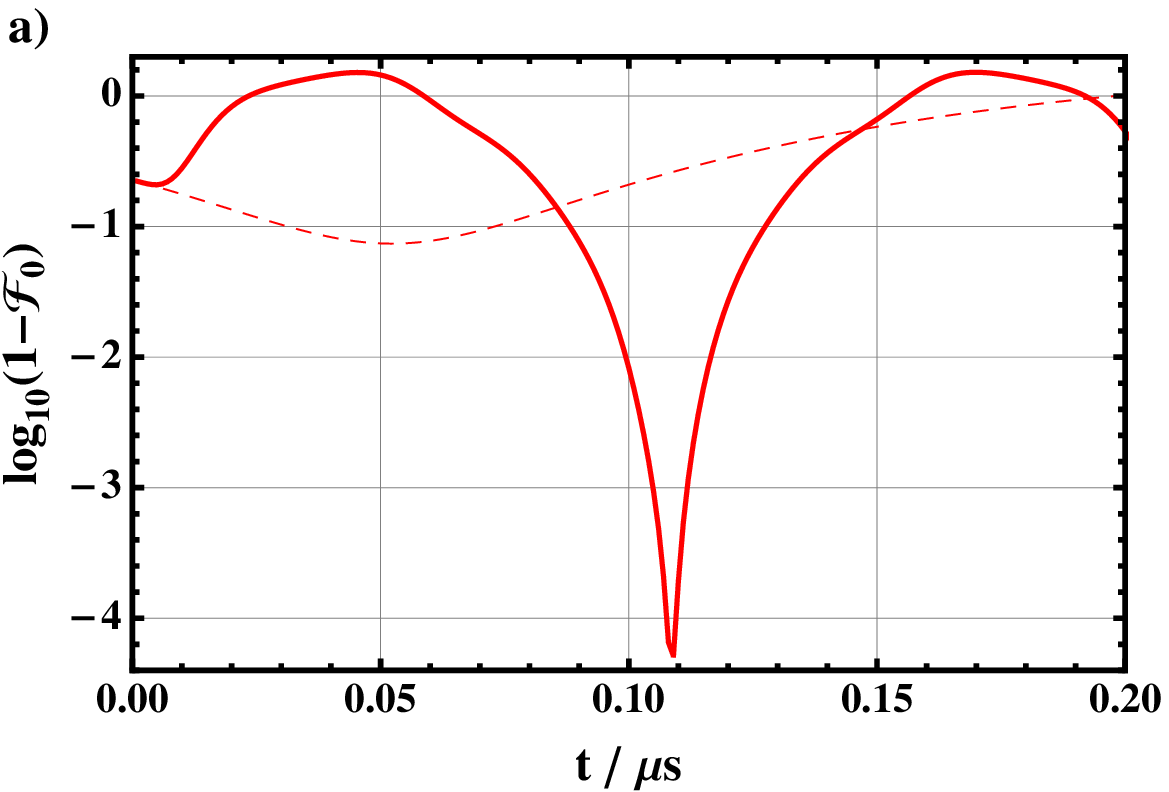}
\includegraphics[width=0.4\textwidth,angle=0]{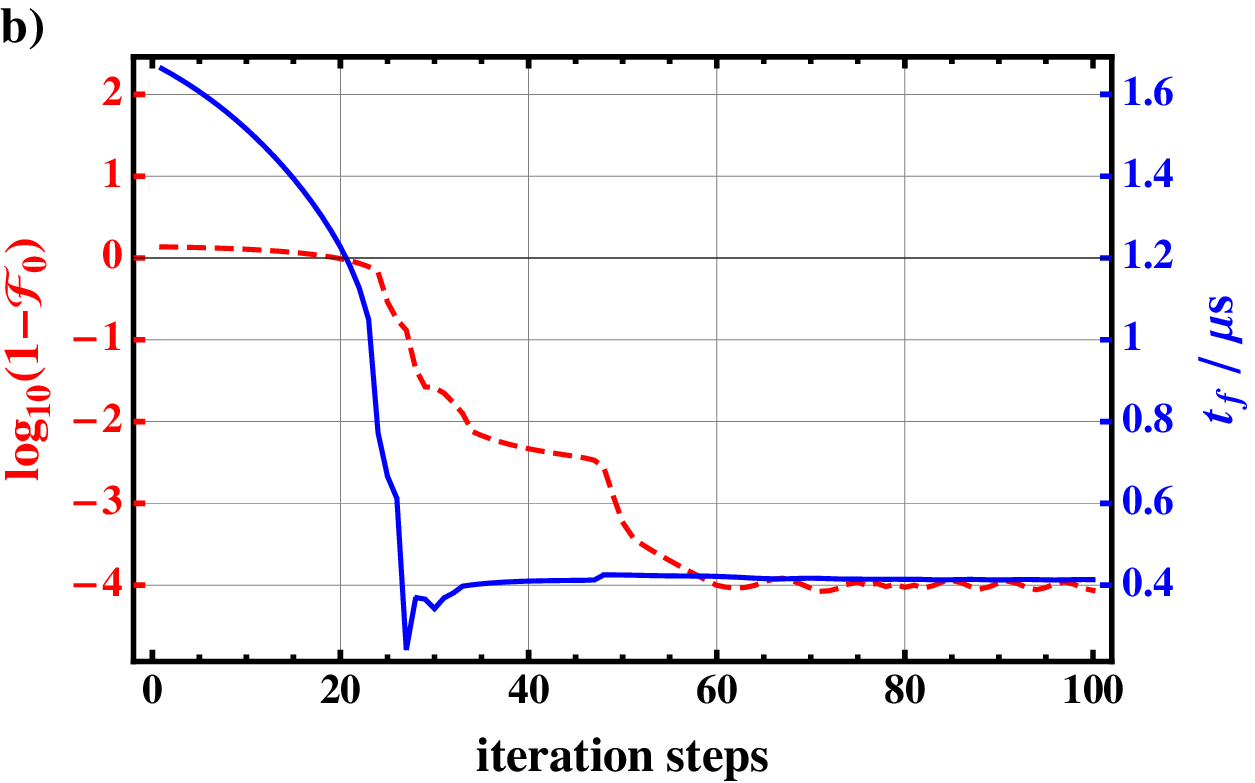}
\caption{a) Logarithmic infidelity as a function of time for the implementation of the gate $\mathcal{U}_{(0.5,0.4,0.3)}$ for $g_x=5.40$ MHz, $g_y=9.95$ MHz, $\omega_1=0.13$ MHz, $\omega_2=0.26$ MHz with (solid line) and without control (dashed line) in minimal time. Without control the gate can be realized with only modest fidelity. b) Typical run of the optimization routine to implement an entangling gate in minimal time. The time $t_f$ (solid line) is minimized, such that an operator fidelity (dashed line) of more than 99.99 \% is reached.}
\label{entgate}
\end{figure}

A general unitary gate $\mathcal{U}_d$ acting on two qubits can be described up to local unitary transformations by only three parameters \cite{PhysRevA.63.062309}:
\be
\mathcal{U}_{(\alpha_x,\alpha_y,\alpha_z)}=\exp\left(-i\sum_{k=x,y,z}\alpha_k\sigma_k^{(1)}\otimes\sigma_k^{(2)}\right)
\ee
In order to test the performance of our algorithm, we implemented time-optimal gates that transform a product state into a maximally entangled state. Such maximally entangling gates are characterized by the inequalities $\alpha_x+\alpha_y\geq\pi/4$ and $\alpha_y+\alpha_z\leq\pi/4$ \cite{PhysRevA.63.062309}. Fig. \ref{entgate} a) shows the time evolution of the fidelity $\mathcal{F}_0$ for the controlled and uncontrolled system in the case where the maximally entangling gate $\mathcal{U}_{(0.5,0.4,0.3)}$ is to be implemented in minimal time with a target fidelity of $\mathcal{F}_\text{thr}=1-10^{-4}$. As one can see, $n_\text{max}=6$ Fourier components are sufficient to achieve a very good result with a fidelity of more than 99.99 \%. More frequency components do not yield substantially shorter pulse durations.

The shortest time scale on which entanglement can be created is given by the intrinsic dynamics of the system. This time scale can be estimated by $t_\text{char}=\pi/(4g_\text{max})$, where $g_\text{max}$ is the largest coupling constant in the Hamiltonian \eqref{ham2s}. In the example of Fig. \ref{entgate} a), we have $t_\text{char}\approx 0.08\,\mu$s. This is comparable to the time scale $t_f=0.11\,\mu$s of our control pulse, which suggests that we have found the time-optimal solution. As a rule of thumb, one can say that control takes the longer the better fidelities one wishes to achieve. However, the need for longer pulse durations scales only moderately with the desired fidelity, {\it e.g.} an increase of the fidelity from $\mathcal{F}_0=1-10^{-4}$ to $\mathcal{F}_0=1-10^{-6}$ only requires less than 2 \% more time.

Concerning the convergence of our algorithm, we see that it is not uncommon that the pulse duration is decreased too much in order to reach the initially specified target fidelity $\mathcal{F}_\text{thr}$ (as it occurs in Fig. \ref{entgate} b) after 27 iteration steps). Nevertheless, in these cases the algorithm still finds the optimal solution by increasing the pulse duration again.

\subsection{Creating and maintaining entanglement}\label{sec:plateau}

While it is relatively easy to transfer a system from a product state $\ket{\Psi(0)}=\ket{\psi_1}\ket{\psi_2}$ to a maximally entangled state $\ket{\Psi_E}$ at a given moment $t_f$ in time by using suitably tailored control pulses, it is often difficult to maintain high entanglement over a long time, since the intrinsic dynamics necessarily induces a decline of entanglement.

In order to avoid a sharp peak in entanglement and thus to maintain it over a longer time interval, one can minimize the curvature $\left\lvert\partial^2\mathcal{E}/\partial t^2\vert_{t=t_f}\right\rvert$ of an entanglement measure $\mathcal{E}$ at the moment $t_f$ in time when entanglement is needed. In this work, we used the tangle $C^2=\left\lvert\bra{\Psi(t_f)}\sigma_y\ket{\Psi ^*(t_f)}\right\rvert^2$ \cite{PhysRevLett.80.2245} to measure entanglement. With our method, there is no need to approximate the derivatives of the propagator $\mathcal{U}$ with respect to time (needed to calculate the curvature of $C^2$) using a method of finite differences, since Floquet theory permits us to compute it analytically as
\be
\frac{\partial^n\mathcal{U}}{\partial t^n}=-\sum_{k,\nu} (\nu\Omega-\varepsilon_k)^n\ovl{\nu}{\chi_k}\bra{\Phi_k(0)}e^{i(\nu\Omega-\varepsilon_k)t}.
\ee
Similar to the problem in Sec. \ref{minentang}, one can set up a target functional $\mathcal{F}=\mathcal{F}_0-\mathcal{F}_p$ with the tangle $\mathcal{F}_0=C^2$ as a measure of entanglement and a penalty functional $\mathcal{F}_p=p\left(\partial^2C^2/\partial t^2\vert_{t=t_f}\right)^2$ with a non-negative parameter $p$ that penalizes a sharp peak of the tangle at $t=t_f$. It turned out that choosing a small value of $p$ (below we used $p=10^{-4}$) for the whole iteration is sufficient to decrease the curvature by several orders of magnitude (practically to zero), while maintaining maximal the tangle at $t=t_f$.

\begin{figure}
\includegraphics[width=0.4\textwidth,angle=0]{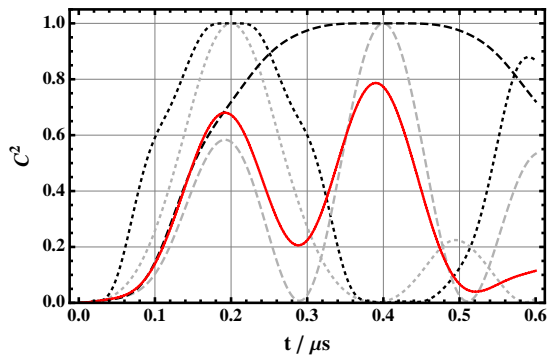}
\caption{Time evolution of the tangle $C^2$ for two interacting spins with $g_x=2.7\,\text{MHz}$, $g_y=6.2\,\text{MHz}$, $\omega_1=0.3\,\text{MHz}$, $\omega_2=0.2\,\text{MHz}$. Without control (solid line) maximal entanglement can never be reached. The grey curves are obtained by maximizing the tangle at $t_f=0.2\,\mu\text{s}$ (dotted curve) and $t_f=0.4\,\mu\text{s}$ (dashed curve), respectively, by using a control pulse with $n_\text{max}=6$ frequency components and fundamental frequency $\Omega=\pi/t_f$. In black are depicted the corresponding curves that one gets if in addition the absolute curvature at $t=t_f$ is minimized. The time interval for which the spins are highly entangled is considerably longer in these cases.}
\label{curv}
\end{figure}

In the following, we consider the system Hamiltonian \eqref{ham2s} with control \eqref{ham2sc} and parametrization \eqref{ham2sp}, the initial product state $\ket{\Psi(0)}$ is chosen at random. In the case displayed in Fig. \ref{curv}, we used the state
\be\ket{\Psi(0)}=\bigotimes_{k=1}^N\left(\cos\left(\theta_k/2\right)\ket{0}+e^{i\phi_k}\sin\left(\theta_k/2\right)\ket{1}\right)\label{instate}\ee
with $\theta_1=1.59$, $\theta_2=2.10$, $\phi_1=5.23$, $\phi_2=0.57$ as the initial state.
As one can see, the curvature of the tangle at the end of the control can be reduced significantly, in the present example from $10^2\,(\mu\text{s})^{-2}$ to $10^{-7}\,(\mu\text{s})^{-2}$, by using a control pulse with only $n_\text{max}=6$ frequency components. In the example for $t_f=0.4\,\mu\text{s}$, the pulse without minimization of the curvature keeps the tangle above 99.9 \% for 4.6 ns, whereas the pulse with minimal curvature guarantees 98 ns of high entanglement, {\it i.e.} more than twenty times longer. The price for this to pay is a moderate increase in the maximal control amplitude from 3.0 MHz to 5.4 MHz. The time interval for which the second derivative vanishes (approximately) - and one has a plateau of high entanglement - is the broader, the longer the duration of the control pulse is. The maximal width $T_\text{max}$ of the plateau of high entanglement is set by the minimal time $t_\text{min}$ that is required to generate maximal entanglement. Since a plateau at $t=t_f$ cannot be extended beyond $t_\text{min}$, the maximal width can be estimated by $T_\text{max}=2(t_f-t_\text{min})$.

It is surprising that the approximation of the tangle to second order in time is valid over such a large interval. Nevertheless, if one increases the pulse duration even further, at some point the time interval for which the evolution of entanglement around $t=t_f$ can be described by the second derivative only reaches its limit. However, in this case one can still enlarge the plateau of high entanglement by maximizing the tangle (and minimizing its curvature) at several moments $t_1,\dots,t_N$ in time. This corresponds to merging several intervals, for each of which the approximation to second order in time is valid. Technically, one has to use the new fidelity $\tilde{\mathcal{F}}=\tfrac{1}{N}\sum_{n=1}^N \mathcal{F}(t_n)$. In the example of Fig. \ref{curv}, the width of the plateau ceases to increase for $t_f\gtrsim 0.6\,\mu$s, but {\it e.g.} by using $\tfrac{1}{2}\left(\mathcal{F}(t_1)+\mathcal{F}(t_2)\right)$ with $t_1=0.6\,\mu$s and $t_2=0.5\,\mu$s as the target, the width of the plateau ($C^2>0.999$) at $t_1=0.6\,\mu$s can be increased by almost a factor of 3.

Interestingly, driving the system from a separable state $\ket{\Psi(0)}$ to a {\it fixed} entangled state $\ket{\Psi_E}$ and then minimizing the curvature has proven very difficult. In this case, the curvature of the fidelity $\lvert\ovl{\Psi(t_f)}{\Psi_E}\rvert^2$ at $t_f$ seems to be fixed by the internal dynamics of the system and the attempt to reduce this curvature always results in a loss of fidelity. This fixed curvature can be explained by the restricted region in Hilbert space that the control can use in order to get from the state $\ket{\Psi(0)}$ to the state $\ket{\Psi_E}$. The constraint of arriving at a precise target state is simply too strong in order to vary a second quantity, in this case the curvature. The use of the tangle as the target functional, on the other hand, allows the control to explore a larger region of the Hilbert space, as there exists a large set of equally entangled states, which can be obtained from the target state $\ket{\Psi_E}$ by local unitary operations.

\subsection{Mediated interaction with inhomogeneous couplings}

\begin{figure}[hb]
\includegraphics[width=0.4\textwidth,angle=0]{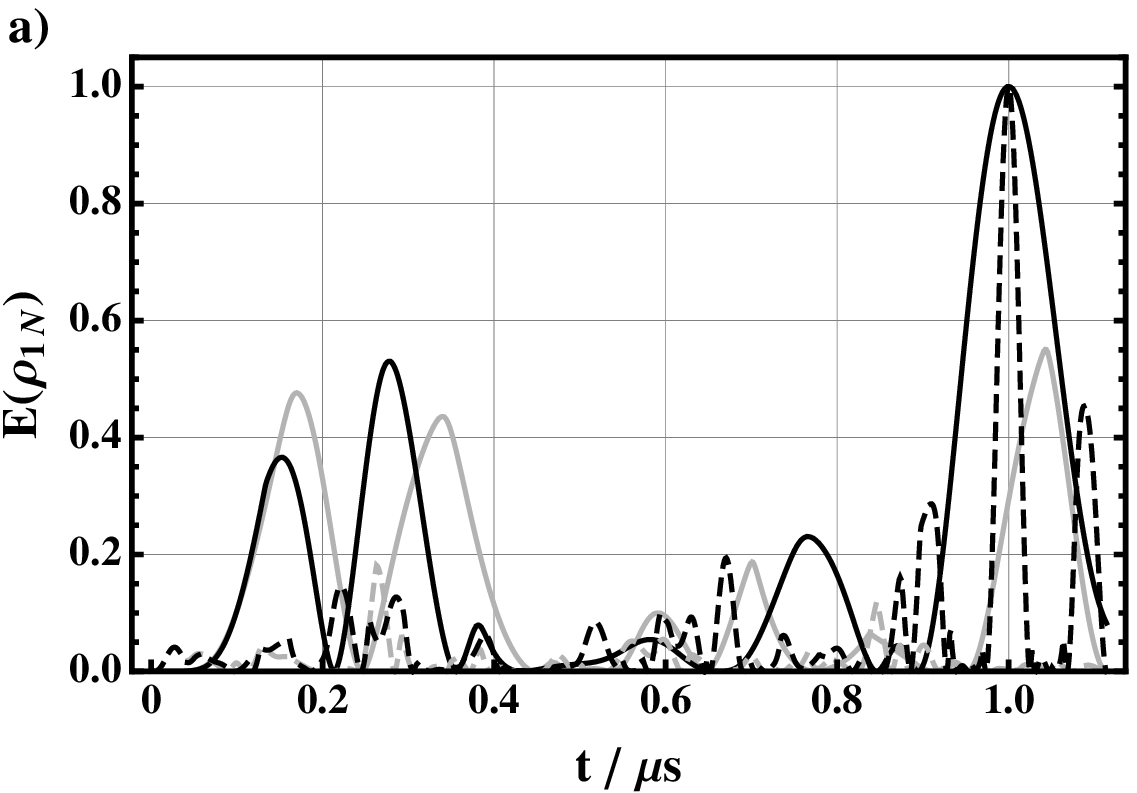}
\includegraphics[width=0.4\textwidth,angle=0]{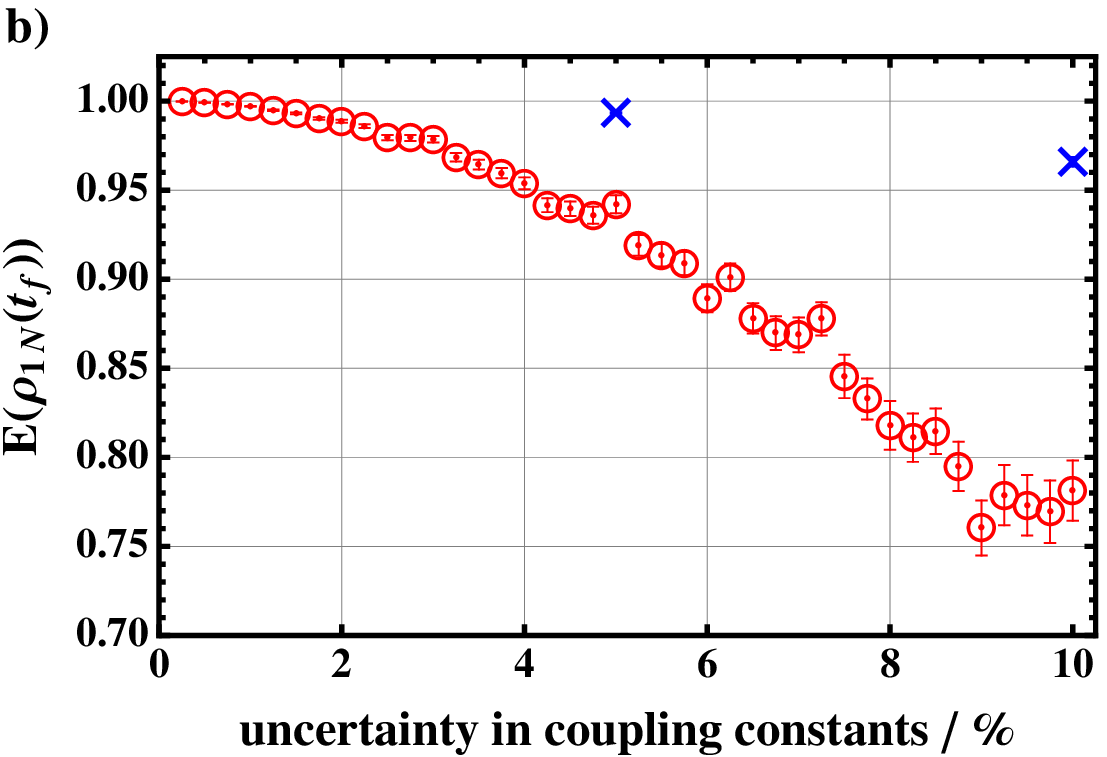}
\caption{a) Time evolution of the entanglement of formation $E$ of the end spins in two random configurations of a chain with $N=3$ (solid line, $g_x^{(1)}=0.78$ MHz, $g_x^{(2)}=1.48$ MHz, $g_y^{(1)}=1.27$ MHz, $g_y^{(2)}=2.65$ MHz, $\omega_1=0.91$ MHz, $\omega_2= 0.97$ MHz, $\omega_3=0.40$ MHz) and $N=4$ spins (dotted line, $g_x^{(1)}=4.36$ MHz, $g_x^{(2)}=1.61$ MHz, $g_x^{(3)}=5.33$ MHz, $g_y^{(1)}=1.02$ MHz, $g_y^{(2)}=8.82$ MHz, $g_y^{(3)}=1.29$ MHz, $\omega_1=0.57$ MHz, $\omega_2=0.55$ MHz, $\omega_3=0.81$ MHz, $\omega_4=0.42$ MHz). In both cases the intrinsic dynamics of the chain (grey) are not able to entangle the end spins, whereas a control pulse with $n_\text{max}=6$ frequency components (black) can generate maximal entanglement. b) Entanglement of formation of the controlled system of $N=3$ spins (circles) if the pulse optimized for the above coupling configuration is applied to a system where the coupling constants are only known up to a certain error. The maximization of the averaged entanglement for a test ensemble of 10 coupling configurations (crosses) significantly increases the amount of entanglement (5 \%: 94.2 \%$\rightarrow$ 99.3 \%; 10 \%: 78.1 \%$\rightarrow$ 96.7 \%).}
\label{chain}
\end{figure}

In solid-state systems the coupling between two spins is often too weak to entangle them, {\it e.g.} if  dipole-dipole interaction is suppressed because of a large spatial separation of the spins \cite{nature.6.912.2010}. Nevertheless, an interaction can still be possible by using the indirect coupling via neighboring spins, as proposed for nitrogen-vacancy centers \cite{PhysRevLett.106.040505,PhysRevLett.110.100503}. With our method we can create entanglement between the end spins of a spin chain, even if the coupling strengths between the spins are only known up to a certain extent. Before discussing pulses that are robust against such variations in the coupling constants, we will first look at the creation of entanglement in the case that all couplings are known exactly:

We assume the Hamiltonian
\be
\mathcal{H}_0 =\sum_{k=1}^N\frac{\omega_k}{2}\sigma_z^{(k)}+\sum_{k=1}^{N-1}\left(g_x^{(k)}\sigma_x^{(k)}\sigma_x^{(k+1)}+g_y^{(k)}\sigma_y^{(k)}\sigma_y^{(k+1)}\right)
\ee
to describe the spin chain and only control the end spins by the same Hamiltonian as in Eq. \eqref{ham2sc}. The goal of entangling the end spins of the chain suggests to use an entanglement measure as the target functional. Unfortunately, the reduced state of the end spins is mixed and most entanglement measures ({\it e.g.} entanglement of formation) are non-analytic for mixed states. This non-analyticity would make the computation of the gradient of the target functional extremely expensive and unreliable. Therefore, we use as the target functional the lower bound \cite{mintert:167902,PhysRevLett.105.020501,Felix2} 
\be
\mathcal{F}=2\tr(\rho_{1N}^2)-\tr(\rho_1^2)-\tr(\rho_N^2)\label{lowb}
\ee
of the tangle of the end spins. Here, $\rho_{1N}$ is the reduced density matrix with respect to the end spins, $\rho_1$ and $\rho_N$ are the reduced single spin density matrices of the end spins. This target functional is only quadratic in the states and permits us to compute all derivatives analytically.

We tested our algorithm on chains of $N=3$ and $N=4$ spins with random couplings $g_{x,y}^{(k)}$ and splittings $\omega_k$ (see Fig. \ref{chain} a)), the separable initial states were also chosen at random ($\theta_1=1.39$, $\theta_2=1.28$, $\theta_3=0.71$, $\phi_1=6.03$, $\phi_2=0.95$, $\phi_3=5.30$ for $N=3$ and $\theta_1=1.60$, $\theta_2=1.31$, $\theta_3=0.94$, $\theta_4=0.44$, $\phi_1=5.24$, $\phi_2=6.21$, $\phi_3=5.85$, $\phi_4=6.07$ for $N=4$, according to the notation \eqref{instate}). In these cases, a pulse with $n_\text{max}=6$ frequency components is sufficient to entangle the end spins with a fidelity of at least $1-10^{-5}$, whereas only very little entanglement is generated without control. 

In contrast to the above case, where the coupling constants were known exactly, experimentally they can only be measured with finite precision \cite{SpinEnv}. In order to estimate the loss of entanglement that results from an uncertainty in the coupling constants $g_{x,y}^{(k)}$, we applied the above pulses on spin chains whose coupling constants deviate by a certain error $\epsilon$ from the original coupling configuration and we calculated the resulting entanglement. The result for $N=3$ is shown in Fig. \ref{chain} b), the error bars come from a test ensemble of 100 different coupling configurations. As one can see, entanglement is deteriorated significantly (by more than 20 \%) if the coupling constants are only known up to 10 \%. A way to avoid this problem is to maximize the target functional not only for one precise coupling configuration but to use as the target functional the averaged fidelity $\langle\mathcal{F}\rangle_\mathbf{g}$ for an ensemble of different coupling configurations $\mathbf{g}=\{g_{x,y}^{(k)}\}_{k=1,\dotsc,N-1}$. This ensemble needs not to be large in order to yield very satisfactory results. As an example, we have optimized the averaged fidelity for an ensemble of only 10 coupling configurations in the case of $N=3$ and then tested it on a larger ensemble of 100 configurations. This method leads to a big increase in entanglement, {\it e.g.} in the case of 10 \% error from 78 to 96 \%. Furthermore, the increase in control amplitude necessary to achieve this robustness is moderate: a factor of 2 for 10 \% error, while for 5 \% even a control amplitude comparable to the one without errors is sufficient.

\section{Conclusions}

Pulse shaping based on variational calculus with Floquet theory yields various advantages that promise to improve our capabilities to control quantum systems.
Beyond the comparatively few parameters to be optimized and the resulting simple, smooth control pulses,
in particular, the analytic dependence of the time-evolution operator on the variable `time' has proven very useful.
As exemplified in Secs. \ref{minentang} and \ref{sec:plateau}, this property easily permits to minimize the duration of gate operations and to extend control targets from single points in time to finite time windows.
This feature can, for example, also be employed to improve the decoupling of a system from its environment. In practice, system-environment interaction with resulting loss of coherence can not be eliminated completely, but a realistically targetable goal is a close-to-optimal revival of coherence after some finite time.
The points in time where revivals occur, however, are changed through the application of control, so that targeting a revival at the point in time when it would have occurred without control may be unnecessarily difficult. Our framework, on the other hand, permits to leave this point in time a dynamical variable that is free to optimization.
Since features like robustness against static noise that are common to typically employed control algorithms \cite{PhysRevA.84.022305,PhysRevLett.111.050404} can also easily be obtained,
the present framework promises to be an attractive option to control quantum systems.

{\bf Acknowledgment}
Stimulating discussion with Tobias N\"{o}bauer, Simeon Sauer, Fabian Bohnet-Waldraff, Andreas Buchleitner and Johannes Majer and financial support by the European Research Council within the project ODYQUENT are gratefully acknowledged.

\bibliography{Referenzen}

\end{document}